\documentclass[9pt, journal]{IEEEtran}

\usepackage{cite}
\usepackage{amsmath,amssymb,amsfonts}
\usepackage{algorithmic}
\usepackage{graphicx}
\usepackage{textcomp}
\def\BibTeX{{\rm B\kern-.05em{\sc i\kern-.025em b}\kern-.08em
    T\kern-.1667em\lower.7ex\hbox{E}\kern-.125emX}}

\begin{document}
\title{\fontsize{16}{18}\selectfont An Ultra-Wideband Leaky Lens Antenna for Broadband Spectroscopic Imaging Applications}
\author{Sebastian H\"{a}hnle, Ozan Yurduseven, \IEEEmembership{Member, IEEE}, Sven van Berkel, \IEEEmembership{Student Member, IEEE}, Nuria Llombart, \IEEEmembership{Senior Member, IEEE}, Juan Bueno, Stephen J. C. Yates, Vignesh Murugesan, David J. Thoen, Andrea Neto \IEEEmembership{Fellow, IEEE} and Jochem J. A. Baselmans \\
\thanks{This work is supported by the ERC COG 648135 MOSAIC. The contribution from Nuria Llombart is supported by ERC Starting Grant ERC-2014-
StG Grant LAA-THz-CC, No. 639749.}
\thanks{S. H\"{a}hnle, J. Bueno, S. J. C. Yates, V. Murugesan and J. J. A. Baselmans are with the Netherlands Institute of Space Research (SRON), Sorbonnelaan 2, Utrecht, 3584 CA Utrecht, The Netherlands (email: s.haehnle@sron.nl, j.bueno@sron.nl, s.yates@sron.nl, v.murugesan@sron.nl, j.baselmans@sron.nl).} 
\thanks{O. Yurduseven is with Huawei Technologies Duesseldorf GmbH, Leprosenweg 1, 82362 Weilheim in Oberbayern, Germany (e-mail: ozan.yurduseven@ieee.orgin).} 
\thanks{S. van Berkel, N. Llombart, D. J. Thoen and A. Neto are with the Microelectronics Department of the Electrical Engineering, Mathemathics and Computer Science Faculty, Delft University of Technology, Mekelweg 4, 2628 CD Delft, The Netherlands (e-mail: s.l.vanberkel@tudelft.nl, n.llombartjuan@tudelft.nl, d.j.thoen@tudelft.nl, a.neto@tudelft.nl).}}

\maketitle

\begin{abstract}
We present the design, fabrication and characterisation of a broadband leaky lens antenna for broadband, spectroscopic imaging applications. The antenna is designed for operation in the 300-900 GHz band. We integrate the antenna directly into an Al-NbTiN hybrid MKID to measure the beam pattern and absolute coupling efficiency at three frequency bands centred around 350, 650 and 850 GHz, covering the full antenna band. We find an aperture efficiency $\eta_{ap} \approx 0.4$ over the whole frequency band,  limited by lens reflections. We find a good match with simulations for both the patterns and efficiency, demonstrating a 1:3 bandwidth in the sub-mm wavelength range for future on-chip spectrometers.

\end{abstract}

\section{Introduction}

The emergence of on-chip spectrometers for far-infrared and sub-millimeter astronomy, such as DESHIMA\cite{Endo2012}, SuperSpec\cite{Shirokoff2014} and Micro-Spec\cite{Cataldo14}, emphasises the need for efficient broadband radiation coupling in the frequency range of $100-1000\ $GHz. These spectrometers offer the ability of measuring medium resolution spectra ($F/\Delta F\sim500$) with an instantaneous relative bandwidth of up to 1:3, complementing high-resolution but bandwidth limited heterodyne instruments.  Future versions of these instruments will move towards multi-pixel focal plane arrays, creating sub-mm imaging spectrometers. These large format focal plane arrays (FPA) are ideally based on broad-band antenna systems with a single beam per feed, allowing a tight sampling of the focal plane. These FPA are typically coupled to reflector systems with large Focal distance to Diameter ratio (F/D) $>$3. Such antennas can also be used as an alternative for CMB missions using multi-color pixels, such as PolarBear which is currently using a sinuous antenna \cite{Inoue2016,Brient13}. 

 We demonstrate the leaky lens antenna as an ideal candidate for these applications. This antenna, first demonstrated by Neto et al. \cite{Neto10_1, Neto10_2} at frequencies up to 70 GHz, is characterised by a 1:3 bandwidth, frequency independent, linearly polarized beams and a high, frequency independent aperture efficiency \cite{Ozan2016}. The first experimental demonstration  at 1.55 THz was presented in 2017 by Bueno et al,\cite{Bueno17}. These experiments demonstrated  super-THz radiation detection with excellent detector sensitivity, but with a limited aperture efficiency $\eta_{ap}=0.24$, evaluated only at a single frequency. The low aperture efficiency was due to a combination of i) radiation loss in the antenna-detector ground plane, ii) a slight misalignment between the feed and the lens and iii) a mismatch, due to fabrication errors, between transmission line and feed. In this paper we demonstrate the fabrication and experimental validation of a leaky lens antenna coupled MKID (Microwave Kinetic Inductance Detector \cite{Day2003}), which does not suffer from these constraints, over a broad frequency band of 300-900 GHz. We measure the absolute coupling efficiency and far field patterns at 3 frequencies: 350, 650 and 850$\ $GHz using narrow band-pass quasioptical filters. From these we obtain an aperture efficiency $\eta_{ap} \approx 0.4$ using a Si lens without anti-reflection coating, which is in good agreement with the predicted performance.


\begin{figure*}[htbp]
\includegraphics{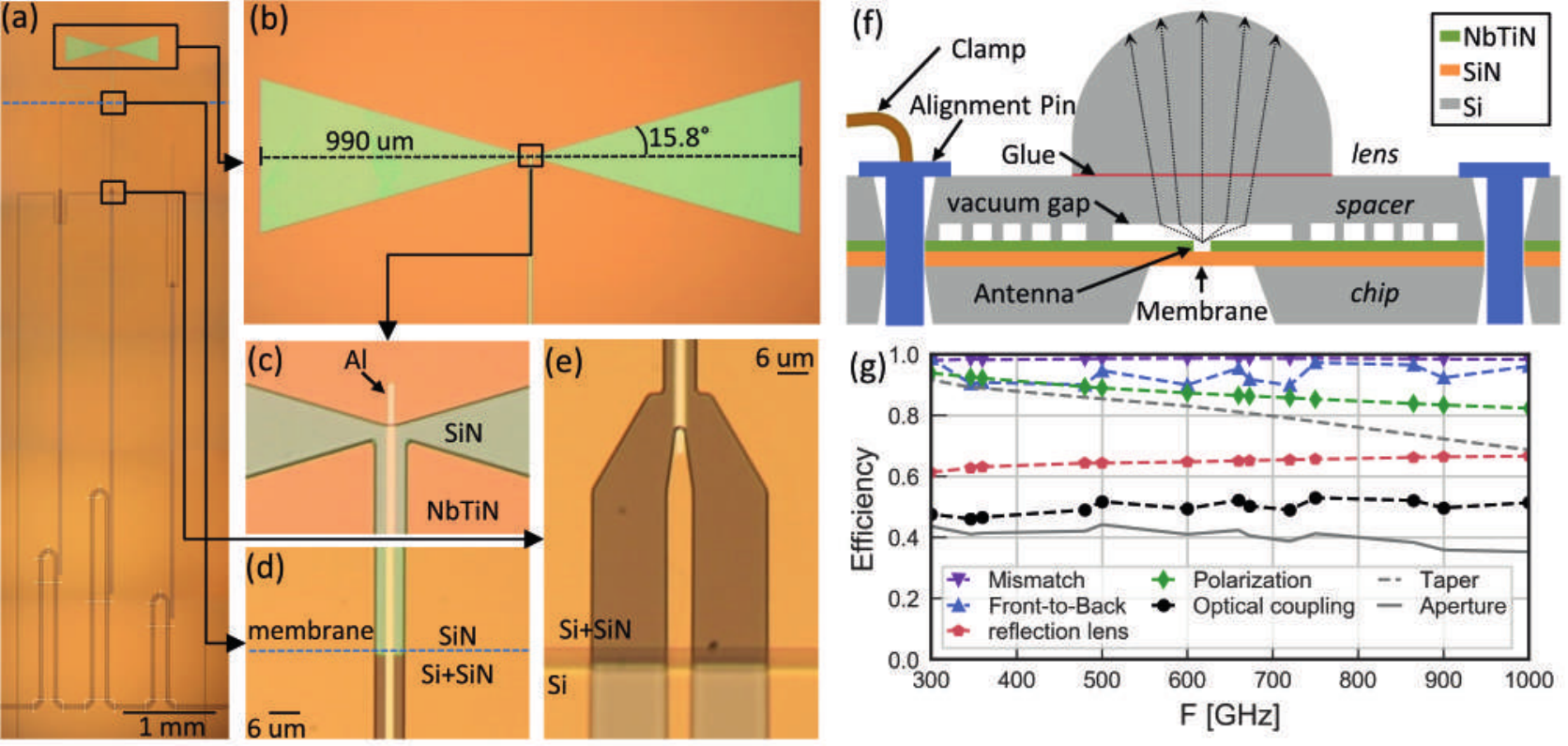}
\caption{(a) Image of the chip center with the tapered antenna slot in the top and the MKID running along the center, coupled to the readout line at the bottom. Additional blind MKIDs are located on the left and right. (b) The leaky wave slot of the antenna on the SiN membrane. (c) Coupling structure to the narrow section of the MKID. The Al center line is shorted via a galvanic connection to the NbTiN ground plane. (d) Narrow MKID section at the membrane edge, marked by the dashed blue line. The $1\ \mu$m thick SiN membrane is transparent and therefore appears slightly blue in the back and front illuminated picture. Below the dashed line, the MKID continues on the SiN layer with Si substrate below. (e) The transition between narrow and wide section of the MKID.
(f) A schematic of the full 3D assembly including the chip, spacer chip and lens. The total stack height is 4.325$\ $mm. The radiation launched in the vacuum gap is refracted at the Si/vacuum interface. (g) The simulated radiation efficiency and its component efficiencies, with $\eta_{rad}= \eta_{mismatch}\eta_{FtB}\eta_{pol}\eta_{reflection}$.} 
\label{fig: device}
\end{figure*}

\section{Device and Antenna Design}
The leaky lens antenna \cite{Neto10_1} is a lens feed consisting of a leaky wave slot on a thin membrane that is  kept at an electrically small distance from the dielectric lens, in order to obtain a directive radiation pattern inside the dielectric which illuminates efficiently the top part of the lens. The leaky wave slot antenna is optimized via simulations in CST with an infinite dielectric medium \cite{Nuria2012}, following the same approach as in \cite{Ozan2016}. The resulting tapered slot combines a high aperture efficiency which is almost constant over a 1:3 bandwidth and is coupled to a single planar feed.  The optimization of the slot to create a frequency independent aperture efficiency results in a tapered slot with a length of $990 \mu$m, slot tapering angle of $15.8^{\circ}$ and a vacuum-gap between antenna and silicon lens of $h = 13.8\ \mu$m. The tapering angle is a compromise between polarization purity and antenna beam width, while the slot length is determined by the lowest frequency. The lens is optimized using a custom Physical Optics (PO) code to create diffraction limited beams with maximized directivity. We use the custom PO code, because the lens size (62 $\lambda$ inside silicon) is too large to allow for full-wave analysis in CST Microwave Studio. This also prevented us from treating multiple reflections inside the lens. This yields a truncated, hyper-hemispherical lens shape with radius of curvature $R = 2.95\ $mm, extension length $L = 1.0\ $mm and truncation angle $\theta = 53.1^{\circ}$ without anti-reflection coating. The diameter of the truncated lens is $5.5\ $mm. We achieve a simulated radiation efficiency $\eta_{rad}$ between $0.45$ and $0.55$ across the whole frequency band as shown in Fig. \ref{fig: device}g. The dominant effect is the reflection efficiency due to the lack of anti-reflection coating at the Si-lens surface. 
Secondly, the polarization efficiency, representing the fraction of the total emitted radiation which is in the co-polarized state, ranges from $94\%$ at low frequencies to $82\%$ at high frequencies due to the relatively large thickness of the vacuum gap which constitutes an inherent tradeoff between polarization purity and directivity. All other contributions to the radiation efficiency, i.e. the feed match ($\eta_{match}$) and front-to-back ratio ($\eta_{FtB}$), are high. The simulated taper efficiency $\eta_{tap} $and aperture efficiency $\eta_{ap} = \eta_{rad}\cdot\eta_{tap}$ are also given in Fig. \ref{fig: device} g. We find that $\eta_{ap}\approx$ 0.4 over the whole frequency band. This is slightly worse than that of corrugated feedhorns, but here we achieve a much larger bandwidth, and comparable to Sinuous antenna's. The development of a broad band anti-reflection coating would increase the aperture efficiency to values close to 0.6.


We couple the leaky wave slot to a narrow coplanar waveguide (CPW) feed with a NbTiN ground and an Al central line (Fig. \ref{fig: device} a-e)). This narrow CPW line is also the last section of an antenna coupled Microwave Kinetic Inductance detector (MKID), which is a $\lambda$/4 CPW resonator with a length of several mm and a resonant frequency of 4.5 GHz. For details on this specific MKID design we refer to Ref.\cite{Janssen13}.  
 Radiation coupled to the antenna (panel b) will propagate into the NbTiN-Aluminium CPW line (panel c), where it will be absorbed only in aluminium central line. This is because the frequency of the radiation (300-900 GHz) exceeds the gap frequency of Al ($F_{\Delta,Al} \approx 93\ $GHz for $T_c \approx 1.28\ $K) but does not exceed the gap frequency of the NbTiN ground plane ($F_{\Delta}\approx 1.1\ $ THz). The NbTiN thereby provides a lossless ground plane for the antenna as well as for the NbTiN-Aluminium CPW line, eliminating ground plane loss as in Ref. \cite{Bueno17}. The radiation absorbed in the aluminium changes the complex surface impedance which results in a slight shift in resonant frequency of the MKID. This shift is detected by a sending a single frequency readout signal at the unperturbed resonant frequency of the MKID through the readout line visible in the bottom of Fig. \ref{fig: device}a. The frequency shift causes a change in transmitted phase in the readout signal that is measured using a homodyne measurement scheme \cite{Day2003}. 
The narrow CPW section has a line width $w = 2\ \mu$m and slot width $2.4\ \mu$m, a thickness of 50 nm and a length of 1.5mm. It is fabricated on top of $1\ \mu$m SiN. Underneath part of this line, the Si is removed to create a free standing membrane, needed for a correct operation of the leaky lens antenna (see Fig.\ref{fig: device}f)). Due to its small width, this section has low radiation loss and a high kinetic inductance fraction, relevant for efficient detection of incoming THz radiation. 
 The remaining resonator therefore consists of a wide CPW of $w = 6\ \mu$m and $s=16\ \mu$m made of NbTiN on bare Si-substrate, with a transition between the sections and step from SiN to Si as shown in Fig. \ref{fig: device}e). The wide section of the MKID then ends in a coupling structure that couples the resonator weakly to the CPW readout line. We choose a very low coupling quality factor of $Q_c = 5000$, because the power absorbed in the device during the beam pattern measurements is $\sim1\ $nW, resulting in a very low internal Q factor of the MKID. Additionally, two meshed layers of low-Tc ($0.65\ $K) $\beta$-phase Ta are located on the front and backside of the chip, to reduce stray radiation inside the chip\cite{Yates2017}. Two blind MKIDs are located close to the antenna-coupled MKID (see Fig. \ref{fig: device}a)) as reference detectors.

\section{Fabrication}
We process the chip from a 375 um thick 4-inch Si wafer coated on both sides with a $1\ \mu$m thick, low tensile stress ($\sim 250\ $MPa) SiN layer, deposited using low pressure chemical vapor deposition (LPCVD) (see Fig \ref{fig: device}f)). The fabrication is similar to the method presented by Bueno \textit{et al.} \cite{Bueno17} with steps as follows: 
i) We etch the SiN on the chip device side with reactive ion etching (RIE) using $35\%$ SF$_6$ and $65\%$ O$_2$ to create a sloped edge for the MKID wide section (Fig. \ref{fig: device}c)). 
ii) We etch the SiN on the wafer backside, to create a hard mask for later Si removal.
iii) We deposit a $350\ $nm thick, low tensile stress NbTiN layer on the device side using reactive sputtering of a NbTi target in a Nitrogen-Argon atmosphere, with the sloped edge of the SiN providing good step coverage for the NbTiN \cite{Thoen2017}. We pattern the NbTiN layer defining the antenna slot, the MKID resonator wide section and narrow section ground plane as well as the microwave readout line. We then etch the NbTiN using the recipe of step (i). 
iv) We sputter a 40 nm thick $\beta$-phase Ta layer on the device side of the wafer, which is then patterned and etched to form an absorbing mesh \cite{Yates2017}.
v) We spin coat $1\ \mu$m polyimide LTC9505 on the wafer, which is patterned and cured into dielectric supports for superconducting bridges along the readout line.
vi) We sputter a $50\ $nm thick Al layer on the device side, which is then patterned and wet etched with a commercially available Al etchant \cite{AlEtch} to define the central line of the narrow MKID section as well as the bridges along the readout line. 
vii) The Si wafer is etched in a KOH bath, with the device side protected with a commercial protection tool, to create the membrane opening.
viii) After the KOH etch, we deposit 40 nm $\beta$-phase Ta on the wafer backside analogous to step (iv). 

To create the 3D structure of the lens-antenna shown in \ref{fig: device}g), we also need a spacer chip and Si-lens. The spacer chip is fabricated on a separate 350 um thick 4-inch Si wafer with SiN on both sides, which is patterned. The wafer is subsequently etched twice in a KOH bath: Once on the lens facing side to create 4 square holes for the alignment pins, which will ensure the alignment between the lens and the antenna, and once on the antenna facing side to create the $14\ \mu$m deep features of the vacuum gap. The Si-lens was made by a commercial partner (Tydex) according to the specifications of our design.

We assemble the device chip, spacer chip and lens in a 2-step procedure: First, the chip and spacer chip are aligned and clamped together using Aluminum alignment pins. The lens is then aligned using a microscope based, home-made tool and glued on top of the spacer wafer using Loctite 406 cyanoacrylate glue, where we use markers on the spacerwafer from the KOH etch for alignment. This glue is chosen for its extremely low viscosity, resulting in a glue gap $<1\ \mu$m.
\begin{figure}
\includegraphics{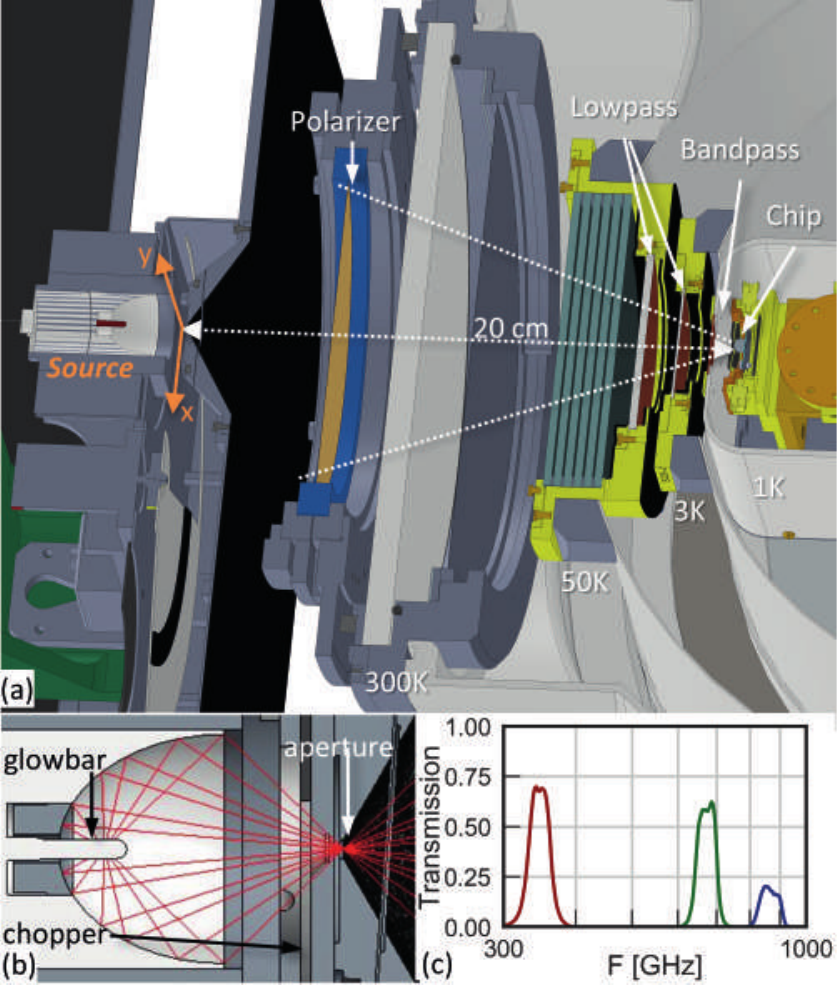}
\caption{ (a) The CAD-model of the cryostat optics and the hot-source in the configuration used for beampattern measurements. (b) A close-up of the source assembly. (c) Filter transmission as function of frequency for the 3 individual filters used in the experiment.}
\label{fig: beampattern_setup}
\end{figure}
\section{Measurements and Results}
\subsection{Beam Patterns}
Measurements of the lens-antenna beam pattern were carried out in the setup shown in Fig. \ref{fig: beampattern_setup}a), where the chip carrying the detector coupled lens-antenna is located on the cold stage of a cryogen free cryocooler equipped with a $^3$He/$^4$He sorption fridge with direct optical access to a source located on an optical table at room temperature \cite{Haehnle2018}. The cryostat reaches a base temperature of $T \sim 265\ $mK with a holdtime of $\sim 32\ $ hours. The opening angle of $\theta = \pm20\deg$ is limited by the rotatable polarizer mounted on the outside of the cryostat's vacuum window. 

The source assembly consists of three main parts: First, a glowbar (Scitec Hawkeye IR-19) operated at $T = 1150^{\circ}$ is used as a multimoded blackbody source, located at one focal point of an elliptical mirror with a small aperture located at the second focus to create a uniform source pattern. We independently validated the source uniformity using a commercial bolometer and rotation stage. Secondly, an eight-bladed chopper, coated with radiation absorber\cite{Klaassen01, Stycast}, is positioned between the two foci. This allows a temperature modulation between the hot source and chopper blade at room temperature. Finally, the glowbar/chopper combination is mounted on an XY-scanner, allowing the measurement of 2D beam patterns of the antenna in the plane of the scanner. Additionally, the cryostat facing side of the source assembly has multiple blackened\cite{Stycast} plates mounted to it, reducing unwanted reflections.

We perform 3 measurements, each with a different band pass filter on the 1K stage of the cooler. The band pass characteristics are shown in Fig \ref{fig: beampattern_setup}c).  They have a sufficiently small bandwidth so that the patterns are not smeared significantly with respect to the step-size of our XY-scanner.
\begin{figure}
\includegraphics{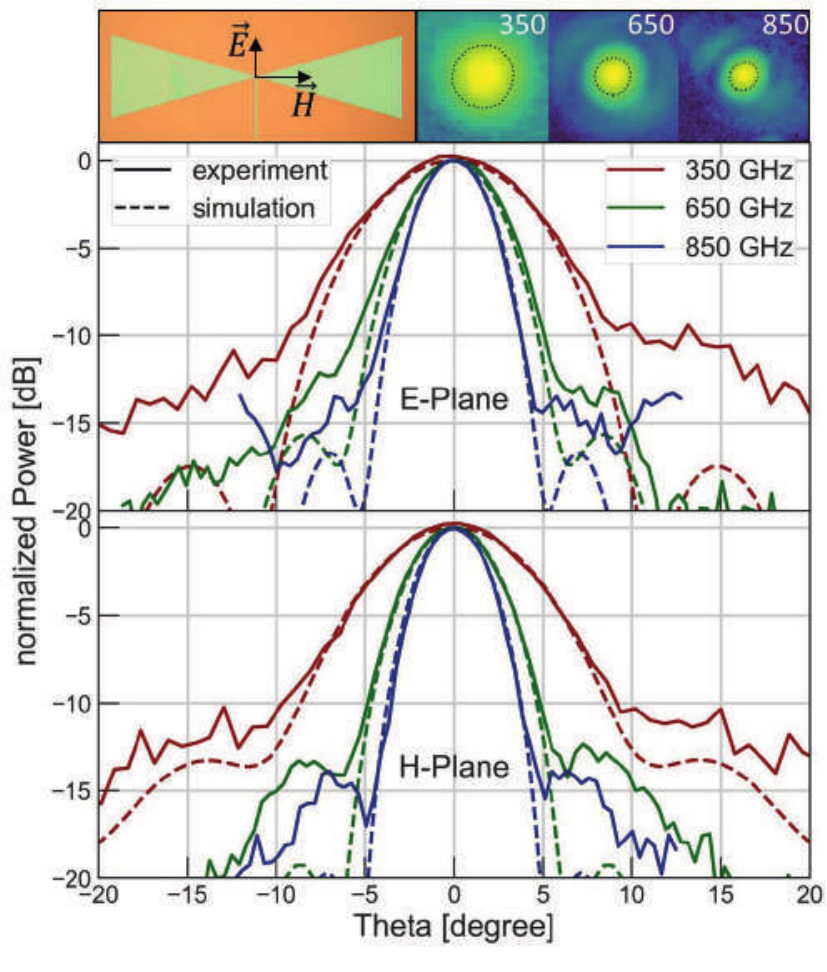}
\caption{The normalized beam patterns in E- and H-plane for all measured frequencies. Insets show $20^{\circ}\times20^{\circ}$ maps of the measured 2D beam patterns (\textit{left} to \textit{right}: $350$, $650$, $850\ $GHz) with a color scale from $-20\ $dB (\textit{dark}) to $0\ $dB (\textit{light}) and a $-3\ $dB contour line. Additionally, the orientation of the antenna slot compared to the measurement planes is shown in the top-left.}
\label{fig: beampattern_result}
\end{figure}
We extract the MKID response as a function of the source position using time-domain Fourier analysis, extracting the peak at the chopper frequency of $f=254\ $Hz for all scanner positions in a step-and-integrate scan mode. The resulting beam patterns are shown in Fig. \ref{fig: beampattern_result} and compared to simulations carried out at the band pass center frequency.
The beam patterns, as simulated using the in-house PO tool, do not take into account multiple reflections that occur inside the lens surface because of the absence of an anti-reflection coating. Multiple reflections can result in a decrease in directivity and an increase in sidelobe-level which is difficult to predict and can change significantly with frequency.


All three frequency bands show excellent agreement between experiment and simulation for the main lobe down to the $-10\ $dB level. The beams are diffraction limited as predicted by simulation, with a full width at half maximum (FWHM) of $3.7^{\circ}$ at $850\ $GHz and $8.3^{\circ}$ at $350\ $GHz  due to the small lens diameter. The leaky-lens antenna is designed for future multi-pixel spectroscopic instruments with closely packed Focal Plane Arrays. Inherent to such designs is that the leaky-lenses are diffraction limited with maximized directivity, as mentioned in Section II. A diffraction limited beam can be described with an Airy pattern with a maximum theoretical Gaussicity of 82\%. The Gaussicity of the beams is about 72\% for all three frequencies and are close to the theoretical maximum.

The measured side lobe levels are around $2$ to $5\ $dB higher than simulated, which we tentatively attribute to multiple reflections inside the lens increasing the effective detected power at large incident angles. An additional possible contribution might be the detection of residual stray-light via surface waves in the chip\cite{Yates2017}. 

\begin{figure}
\includegraphics{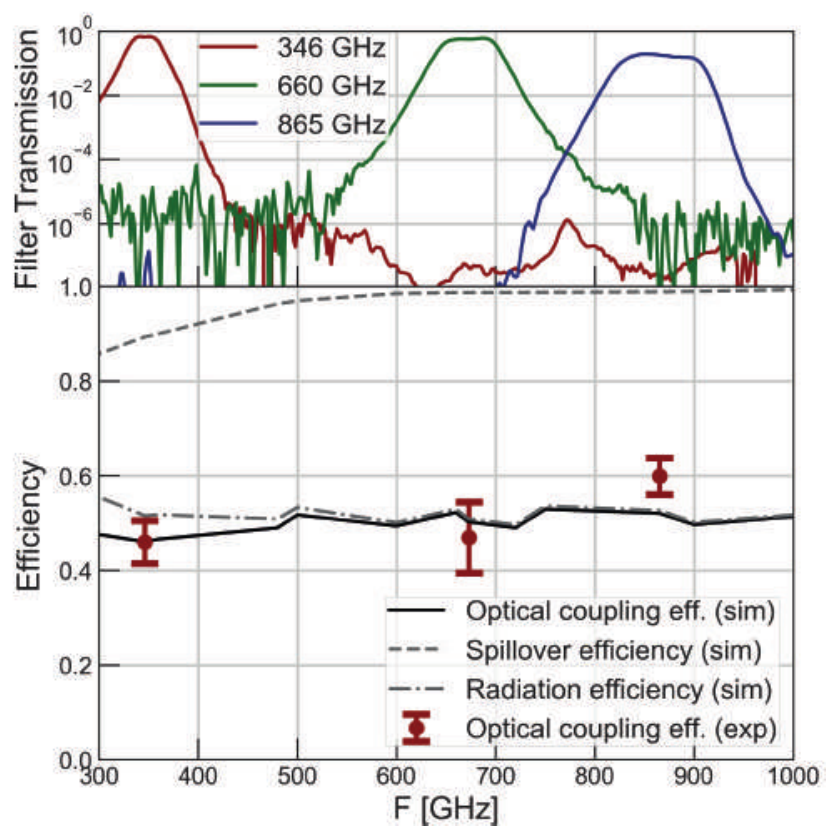}
\caption{ (a) The transmission curves for the three optical filterstacks; (b) The measured optical coupling efficiency, together with the simulated values of the optical coupling efficiency and its contributions, the radiation efficiency and spillover efficiency between the lens-antenna and the calibration source.
}
\label{fig: efficiency_result}
\end{figure}

We measure the optical efficiency in a separate cryostat which provides a highly controlled environment and is described in detail in Ref. \cite{Baselmans2012}. For this experiment the chip is mounted in a light-tight box on the cryostats cold stage, an adiabatic dilution refrigerator (ADR), operated at a stable $120\ $mK. An opening in the light-tight box provides a field-of-view through multiple optical filters to a thermal calibration source with a temperature range of $2.7\ $K - $40\ $K. The optical filterstacks correspond to the bands of the beam pattern measurements (see Fig. \ref{fig: efficiency_result}). However, their out-of-band rejection is much larger, which is needed to sufficiently reject low frequency radiation from the black body source operating at low temperatures.

The angular throughput to the source is limited by an aperture plate to an angle of $\pm10^{\circ}$. This is large enough to couple almost to the whole beam at $350\ $GHz, but avoids straylight from large angles. 

\subsection{Optical Efficiency}
We determine the optical  coupling efficiency $\eta_{opt}(\nu_0)$ of the MKID coupled lens-antenna at a frequency $\nu_0$ experimentally by measuring the noise-equivalent power $NEP_{exp}$ at a source temperature $T_s$: 
\begin{equation}
\eta_{opt}(\nu_0) = \frac{\int2P_{s,\nu}h\nu d\nu + \int\frac{4\Delta P_{s,\nu}}{\eta_{pb}}d\nu}{NEP_{exp}^2 - \int2P_{s,\nu}h\nu F_{\nu}O_{\nu}d\nu}
\end{equation}
with the superconductor bandgap $\Delta$, the pair-breaking efficiency $\eta_{pb} \sim 0.4$ for our film \cite{Guruswamy}, the filter transmission $F_\nu$, occupation number $O_\nu$ and the single-moded source power after filter transmission $P_{s,\nu}$. A detailed description and derivation of this equation and its components can be found in Ferrari \textit{et al.} \cite{Ferrari18}.

This method is valid for narrow frequency bands and photon-noise limited MKIDs, and allows us to obtain precise absolute measurements of the optical efficiency at multiple frequencies across the whole band of the lens-antenna. For our Al based hybrid MKIDs, photon noise limited performance is proven experimentally, if the noise spectrum is white with a roll-off consistent with the quasiparticle lifetime. This is true for the detector presented at absorbed power levels $P>10\ $fW. 
While it is in principle possible to measure the response over the whole band using a Fourier-transform spectrometer (FTS) \cite{Bueno17}, this only provides a relative power measurement and not an absolute validation of the optical coupling efficiency of the detector .

The optical coupling efficiency measurements are carried out at multiple  $T_s$, corresponding to a power range between $10\ $fW and $1\ $pW, which are averaged to obtain the values shown in Fig. \ref{fig: efficiency_result} in comparison with simulation. for a detailed description of the experimental procedure we refer to Ref.\cite{Janssen13}. As shown in bottom panel of Fig. \ref{fig: efficiency_result} we find excellent agreement between the measured and simulated optical coupling efficiency $\eta_{opt} = \eta_{rad}\cdot \eta_{Spillover}$ for all frequency bands. Here $\eta_{spillover}$ describes the spillover  between the lens-antenna beam and the source, which is close to unity except for the lowest frequency. Hence this result demonstrates that the measured $\eta_{rad}$ is in good agreement with our model calculation. Together with the good agreement of the patterns which validate the model calculation of $\eta_{tap}$, we can conclude that the aperture efficiency $\eta_{ap} = \eta_{rad}\cdot\eta_{tap}$ is well described by our model, and given by $\eta_{ap}\approx 0.4$ over the whole frequency band (see Fig.\ref{fig: device}). Only at $850\ $GHz we see a slightly higher coupling, possibly due to a small stray light contribution: The beam width is much smaller than the source aperture at this frequency. 

\section{Conclusion}
In conclusion, we have measured beam patterns and optical efficiency of a leaky lens antenna over a 1:3 bandwidth around $350\ $GHz, $650\ $GHz and $850\ $GHz. We  infer an aperture efficiency of the lens-antenna $\eta_{ap}$ = 0.4 over the whole frequency band, based on measurements of the absolute coupling efficiency between the detector and a thermal calibration source together with a good agreement between measured and simulated beam patterns. The aperture efficiency is limited primarily by the antenna radiation efficiency, which is low due to the absence of an anti-reflection coating. The agreement between simulation and experiment show the validity of the design process, allowing for the design of optimised lens-antennas for on-chip spectrometer or CMB applications. 
The development of a broadband anti-reflection coating would allow for a significant increase in optical efficiency by up to 30\%.f Given the wide angular extent of the primary field in the dielectric an anti-reflection coating based upon meta-materials is extremely difficult to machine due to the steepness of the lens. An interesting alternative would be the use of stacks of foam as proposed by Ref. \cite{Nadolski} or the use of thermal spray coatings \cite{Inoue2016}, which allow for a more practical approach towards a conformal coating. Changing the lens geometry and feed position could allow for frequency stable patterns\cite{Filipovic93}, with lower directivity, which would allow for a more efficient broad-band coupling to a reflector system.

\bibliography{FINAL_VERSION}
\bibliographystyle{IEEEtran}

\end{document}